\documentclass[nobm,nocite]{epl}
\newcommand{\BEQ}{\begin{equation}}
\newcommand{\EEQ}{\end{equation}}
\newcommand{\BEA}{\begin{eqnarray}}
\newcommand{\EEA}{\end{eqnarray}}

\renewcommand{\d}{{\rm d}}

\newcommand{\oq} {\frac{1}{4}\,}
\newcommand{\os} {\frac{1}{6}\,}

\newcommand{\mt} {{\bf m}}
\newcommand{\eps}{\,\epsilon\,}

\newcommand{\va} {{v_b}}

\newcommand{\qb} {{\bf q}}

\newcommand{\half}{\frac{1}{2}\,}
\newcommand{\bd}{{0}}
\newcommand{\ub}{{ub}}

\def\simg{\mathrel{\mathchoice {\vcenter{\offinterlineskip\halign{\hfil
$\displaystyle##$\hfil\cr>\cr\sim\cr}}}
{\vcenter{\offinterlineskip\halign{\hfil$\textstyle##$\hfil\cr
>\cr\sim\cr}}}
{\vcenter{\offinterlineskip\halign{\hfil$\scriptstyle##$\hfil\cr
>\cr\sim\cr}}}
{\vcenter{\offinterlineskip\halign{\hfil$\scriptscriptstyle##$\hfil\cr
>\cr\sim\cr}}}}}

\begin{document}
\title{Walks of molecular motors in two and three dimensions}
\shorttitle{Walks of molecular motors in $2$ and $3$ dimensions}

\author{Theo M. Nieuwenhuizen\inst{1,2}, Stefan Klumpp\inst{2},
   \and Reinhard Lipowsky\inst{2}} \shortauthor{Theo Nieuwenhuizen \etal}
\institute{ \inst{1} Institute for Theoretical Physics,
   University of Amsterdam,\\
   Valckenierstraat 65, 1018 XE Amsterdam, The Netherlands\\
   \inst{2} Max-Planck Institute for Colloid- and Interface Research,
   \\ MPI KG, 14424 Potsdam, Germany } \date{printout \today; version
   February 8, 2001}


\maketitle

\begin{abstract}
   Molecular motors interacting with cytoskeletal filaments undergo
   peculiar random walks consisting of alternating sequences of
   directed movements along the filaments and diffusive motion in the
   surrounding solution.  An ensemble of   motors is studied
which interacts with a single filament in two and three dimensions.
The time evolution of the probability distribution for the bound and
unbound motors is determined analytically. The diffusion of the motors
is strongly enhanced parallel to the filament.
   The  analytical expressions are in excellent agreement with the results of
   Monte Carlo simulations.
\end{abstract}

\section{Introduction}

Small particles,  dispersed
in a liquid, constantly undergo random movements. This was first
observed by the botanist Brown and later explained by Einstein and
Smoluchowski in terms of thermally--excited collisions. The 
corresponding diffusion
coefficient depends on the temperature $T$ and on the viscosity of the
liquid which reflects the balance between thermal fluctuations and
viscous dissipation.  Such diffusive processes are also ubiquitous in
biological systems \cite{berg93} and, in particular, within biological
cells \cite{luby00}.

When the diffusing particles are attracted towards a 2--dimensional surface or
a 1--dimen\-sional filament, they can attain a bound state in which 
their motion is
restricted to this low--dimensional manifold.  Even though the 
corresponding diffusion
coefficient is expected to be small, the reduced dimensionality of the
diffusion process can enhance the probability for two diffusing
particles to collide and, thus, the rate of diffusion--controlled
reactions \cite{adam68}.

Molecular motors can bind to filaments such as   microtubules or DNA 
and then undergo
1--dimensional  diffusion or 'sliding' in the absence of adenosine triphosphate
(ATP).  This has been observed both for cytoskeletal motors at microtubules
\cite{vale89} and for RNA polymerase at DNA \cite{guth99}. In addition,
these motors are    able to hydrolyse ATP and to use the released
free energy in order to change their own conformation.
\cite{krei99,howa01} In this way, detailed balance
is broken and the molecular motors can perform directed walks along the
filaments as studied in the framework of ratchet models, see, e.g.,
\cite{lipo171} and the extensive review in \cite{reim02}.

Since the binding energy, $\Delta E$, between the motor particle and
the filament is necessarily finite, the motor particle will again
unbind from the filament after a certain walking time, $\Delta t_{b}$,
which is of the order of $ \exp[\Delta E/k_B T]$.  On time scales
which are large compared to $\Delta t_{b}$, the motor particle
undergoes peculiar random walks which consist of alternating sequences
of directed movement and unbiased diffusion \cite{ajda95,lipo173}.  Such
walks can be studied in the context of
  lattice models  which we have recently introduced and studied
for open and closed compartments using scaling arguments and
Monte Carlo (MC) simulations \cite{lipo173}.

In the present work, we consider   unbounded
geometries, i.e.,   systems without confining walls, for which we are
able to obtain {\em analytical} solutions both in two and in
three dimensions. Furthermore, we show that the
  {\em diffusion  of both the bound and the unbound motors} is 
strongly enhanced parallel
to the filament. For large times,
the corresponding diffusion coefficients are found to attain    anomalously
large  values  in two dimensions and to exhibit   large
logarithmic correction terms in  three dimensions.
All analytical results are compared with and confirmed by extensive
MC simulations as shown in  Fig.~\ref{f.1} and \ref{f.2} below.

\section{2-dimensional case ($d =2$)}
Consider a discrete time random walk on a square lattice with lattice
sites labeled by integer coordinates $(n,m)$.  At each step,  a motor 
'particle'
  has probability 1/4 to jump to any of its four nearest neighbor
sites.  On the line with  $m=0$ which corresponds to the filament, the motion
is different:  The particle jumps forward from
$(n,0)$ to $ (n+1,0)$ with probability
$1-\gamma-\frac{1}{2}\delta-\frac{1}{2}\epsilon$
and backward from $(n,0)$ to $(n-1,0)$  with probability $\half 
\delta$, where $\gamma$
is the probability to make no jump at all.  The latter parameter is included in
order to incorporate the large difference between  the diffusion 
coefficient in the
solution and on the filament \cite{lipo173}.  In addition, the 
particle leaves the
filament and, thus,  jumps from $(n,0)$ to  $(n,\pm 1)$ with probability
$\frac{1}{4}\epsilon$ where the parameter $\epsilon$ is taken to be small.

As long as the motor  is bound to
  the filament with $m=0$,  it has the  average velocity $\va
=1-\gamma-\delta-\half\, \epsilon$. It also has the
probability $ \half\, \eps $ to unbind per unit time.  Thus, the 
probability that the motor
is still bound after $t$ time steps is   $(1-\half\eps)^t\approx
\exp(-\half \eps\, t)$.

As the initial distribution,  we take an ensemble of non-interacting particles
at $n=m=0$.  The master equation for this dynamics can be solved using
Fourier--Laplace (FL)  transforms, see, e.g.,  \cite{Weiss}.
  The FL transforms for the full probability  $P$  and
for the probability   $P_b$  to be bound to the filament are given by
\BEQ
P(q,r,s)=\sum_{t=0}^\infty\sum_{m,n=-\infty}^\infty \frac{e^{iq m+ir
     n}}{(1+s)^{t+1}} P_{n,m}(t) \quad {\rm and} \quad
P_b(r,s)=\sum_{t=0}^\infty\sum_{n=-\infty}^\infty \frac{e^{ir
     n}}{(1+s)^{t+1}} P_{n,0}(t)
\label{DistributionsDefined}
\EEQ
and can be related to each other due
to the translational invariance of the system parallel to the 
filament. By integration over
$q$,  one then obtains  the explicit solution
\BEA
  P_b(r,s) &=&\frac{1}{s+(1-\gamma)(1-\cos r)
   +\half\eps(\cos r-e^{-\mu})-i\va\sin r}
\label{FLTDistributionBoundMotors}
\EEA
where $\cosh\mu=2+2s-\cos r$.  This result is easily checked for
$\epsilon=0$ (random walk in one dimension) and for $\gamma=0$,
$\delta=\half$, and $\epsilon=1$ (non-biased random walk in two dimensions).
Likewise, one obtains a closed expression for    $P(q,r,s)$
and for the   FL transformed distribution of the unbound
motors  which we define via  $P_{ub}(q,r,s) \equiv P(q,r,s) - P_b(r,s)   $.

\section{Fraction of bound motors}
The   fraction $N_0(t)   \equiv \sum_n P_{n, 0}(t)$
of  motors bound to the filament  can be calculated in closed form
which leads to

\begin{figure}[t]
   \onefigure[scale=0.29]{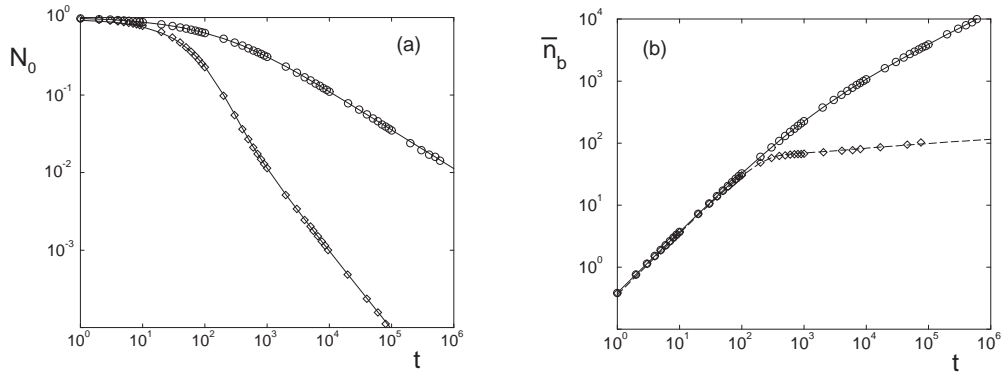}
\caption{(a) Fraction $N_0$  and (b) Average position ${\overline n}_b$
of  motors bound to the filament as a function of
time $t$. In both figures, the circles
and diamonds are  MC data for $d=2$ and $d=3$,  respectively,
and the curves represent the analytical results, see text.
  For large $t$, the bound fraction $N_0 \sim 1/\sqrt{t}$ in $d=2$ (circles) and
$N_0 \sim 1/t$ in   $d=3$ (diamonds).
The parameters, which determine the jump probabilities of the 
underlying random walks,
have the values  $\gamma =0$, $\delta = 0.6$, and  $\epsilon = 0.05$. }
\label{f.1}
\end{figure}

\BEQ
  N_0(t)=\oint\frac{\d s}{2\pi i}(1+s)^{t}\,P_b(r=0,s)
=\int_0^{1/\eps^{2}}\frac{\d y}{\pi\sqrt{y}}\,\frac{(1-\eps^2
   y)^{t+1/2}} {1+(1-2\eps)y}\approx \int_0^\infty\frac{\d
   y}{\pi\sqrt{y}}\,\,\frac{e^{-y\epsilon^2t}} {1+y}
\EEQ
where the asymptotic equality  holds for small $\eps$ and large
$t$.  For small $t$,  the survival fraction $N_0(t)$ can be expanded in powers
of  $\epsilon\sqrt t$ which leads to
$N_0(t) \approx 1-2\,\frac{\epsilon \sqrt t}{\sqrt\pi}+\epsilon^2t
-\frac{4\epsilon^3t^{3/2}}{3\sqrt\pi}$.
Thus, although the motors detach at times   $\sim
1/\eps$, they stay close to the filament until times  $t \sim 1/\epsilon^2$.
For $t\gg 1/\epsilon^2$, on the other hand,  the survival fraction decays
to zero as  $N_0(t) \approx
(1-\frac{1}{2\epsilon^2t} 
+\frac{3}{4\epsilon^4t^2})/\sqrt{\pi\,\epsilon^2\,t}$.
The complete time evolution of  $N_0(t)$,
as obtained by numerical evaluation of the exact integral in (3),  is 
displayed in
Fig.~\ref{f.1}(a).  We have also determined  this quantity in a 
completely different way
using Monte
Carlo (MC) simulations.  Inspection of Fig.~\ref{f.1}(a) shows that 
the results of these
two  approaches agree very well.

\section{Average position on the filament}
The first moment of the probability distribution $P_{n,0}$ for the   motors
on  the filament   is given by
\BEQ
  \label{N1t=i}
N_1(t) \equiv \sum_{n}n \, P_{n,0}(t) =\frac{2\va}{\pi \epsilon^2}\int_0^\infty
\frac{\d y}{\sqrt{y}} \,\frac{1-e^{-y\epsilon^2t}}{(1+y)^2}
\EEQ
as follows from an expansion of (2) to first order in $r$.
For small $t$, this leads to $ N_1(t) \approx \va 
t\,[1-(8/3{\sqrt\pi}){\epsilon
  \sqrt t}+\cdots]$.  In the same limit, the  average position
${\overline n}_b \equiv N_1/N_0$ and the average
velocity  $ {\overline v}_b \equiv \d\,{\overline n}_\bd/\d t $   behave    as
${\overline  n}_b \approx \va t(1-\frac{2}{3}\,\epsilon \sqrt t/\sqrt\pi)$
and    $ {\overline v}_b \approx\va (1-\frac{ \epsilon \sqrt 
t}{\sqrt\pi})$, respectively.
For large $t$,  the integral in (\ref{N1t=i}) leads to
$N_1(t) \approx \frac{\va }{\epsilon^2}(1-\frac{2}{\epsilon\sqrt{\pi t}})$ and
thus to

\BEQ
  {\overline n}_b  \approx  \frac{\va
   \sqrt{\pi t}}{\epsilon} (1-\frac{2}{\epsilon\sqrt{\pi t}})  \qquad
{\rm and } \qquad
   {\overline v}_b \approx  \frac{\pi}{2}N_0(t)\,\va \approx
   \frac{\sqrt{\pi}}{2} \frac{\va}{\eps \sqrt{t}} \quad .
\EEQ
  The complete time evolution of the   average
position $ {\overline n}_b $
is displayed in Fig.~\ref{f.1}(b). Again, we compare $ {\overline n}_b $ as
  obtained   from  the
integrals in (3) and (4) with MC data and find very good agreement.
Note that the average position ${\overline n}_b$ of the bound motors behaves
as $\sim \sqrt{t}$ for large $t$.  As shown below, the same time dependence
applies to the average position of {\em all} motors    as previously
predicted by   scaling arguments \cite{ajda95,lipo173}

\section{Variance and diffusion coefficient on the filament}
\begin{figure}[t]
   \onefigure[scale=0.3]{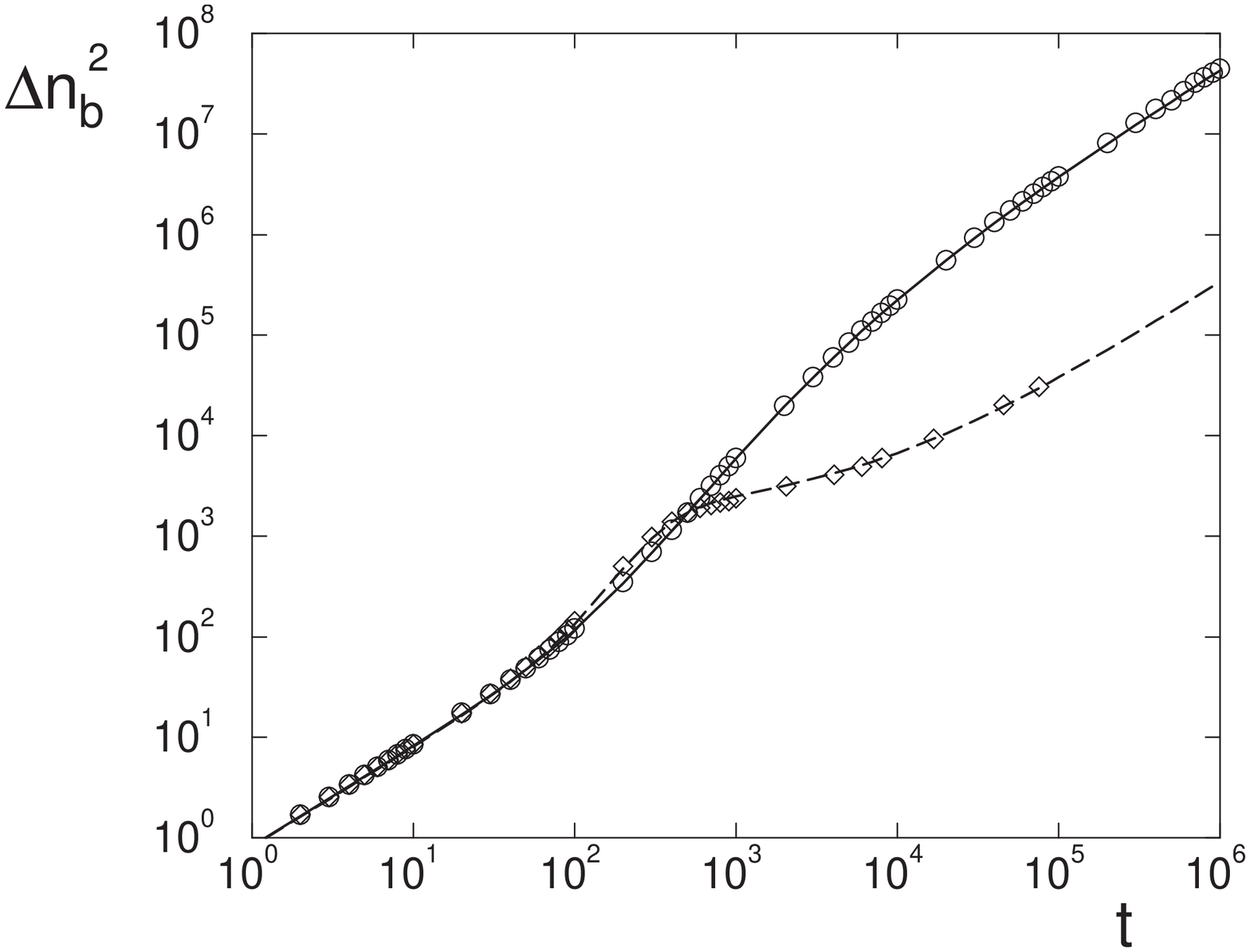}
\caption{Positional variance $\Delta n^2_b$ of bound motors  as a 
function of time
$t$.  The MC data for $d= 2$ (cirles) and $d =3$ (diamonds)   are 
again compared
with the analytical results  (full and dashed curve).
For large $t$, the   diffusion
coefficient $ D_b (t)  = \half  \d \Delta  n^2_b / \d t $ attains the
anomalously large value
$D_b (\infty) = \frac{\va ^2}{2\epsilon^2} (4-\pi)$ in $d=2$ (circles)
whereas    $D_b (t) \approx \frac{1}{6} +  \frac{9 \va^2 }{2 \pi^2 \epsilon^2}
\frac{\ln t}{ t}$ in $d=3$ (diamonds)  with the usual diffusion constant
$D_b (\infty) = \frac{1}{6}$ but a large correction
term $\sim \ln t/ t$. The jump probabilities have the same values as in
Fig.~\ref{f.1}.}
\label{f.2}
\end{figure}

The second moment of the probability distribution $ P_{n,0}$ for the 
bound  motors
    is
\BEQ
  N_2(t) \equiv \sum_{n} n^2 \, P_{n,0}(t) \approx \frac{1-\gamma}{\va}N_1(t) +
  \frac{2\va ^2}{\pi\epsilon^4}\int_0^\infty\d x \,
\frac{(1-e^{-x\epsilon^2 t})(1-3x)}{(1+x)^3x^{3/2}}
\EEQ
as obtained from an expansion of  (\ref{FLTDistributionBoundMotors})
to second order in $r$
and for small $\eps$.
For small $t$, this leads to $N_2(t) \approx (1-\gamma)t   + \va ^2t^2
(1-\frac{16}{5}\,\frac{\epsilon \sqrt t}{\sqrt\pi})  $. The
   positional variance,   defined via
$ \Delta n^2_b \equiv \frac{N_2(t)}{N_0(t)}
-\frac{N_1^2(t)}{N_0^2(t)} $, i.e.,  with respect to
the conditional probability $ P_{n,0}(t) / N_0(t)$, then behaves as
\BEQ
  \Delta n^2_ b  \approx (1-\gamma)t +
\frac{2\va^2}{15}\,\frac{\epsilon\,t^{5/2}}{\sqrt\pi}
\quad {\rm for } \, \, {\rm small} \, \, t \quad .
\EEQ
The relative variance ${ \Delta n^2_b}/{\overline{n}^2_b}
\approx \frac{2}{15}\,\frac{\epsilon\,\sqrt t}{\sqrt\pi}$ is small
since $t\ll 1/\epsilon^2$.
We now  define a {\em time-dependent} diffusion
coefficient via $ D_b (t)  \equiv \half  \d \Delta
   n^2_b / \d t $ which exhibits the small--$t$ behavior
\BEQ
  D_b (t) \approx \half(1-\gamma)+\frac{1}{6\sqrt\pi}\,\va
^2\,\epsilon t^{3/2} \quad .
\label{DboundSmallt}
\EEQ
In addition, the behavior for large $t$ is found to be
  \BEQ
\Delta n^2_b
\approx \frac{\va ^2}{\epsilon^2} (4-\pi-\frac{2\sqrt\pi}{\epsilon\sqrt
   t})\,t \quad  {\rm and } \quad
   D_b (t)  \approx \frac{\va ^2}{2\epsilon^2}
(4-\pi-\frac{\sqrt\pi}{\epsilon\sqrt t})
\quad .
\EEQ
Note that the diffusion constant $  D_b (\infty) \sim v_b^2 / 2 \epsilon^2$
is quite large compared to the value $D_b(0) = \half(1-\gamma)$ as appropriate
for the 1--dimensional diffusion of a bound motor.  In fact,
  one has  $D_b(t) \sim v_b^2 /   \epsilon^2 $ as soon as
  $t \simg 1/\epsilon^2$, see (\ref{DboundSmallt}).
This broadening occurs since the unbound motors lag behind
the bound ones.
The time--dependence of  $\Delta n^2_b $ is shown in Fig.~\ref{f.2}.

\section{Probability distribution on the filament}
At large times,  $N_2/N_0 \sim (N_1/N_0)^2$ which indicates that the
probability distribution attains a scaling form. It follows from
(\ref{FLTDistributionBoundMotors})
that $P_b (r,s) \approx {1}/{(s-i\va
   r+\epsilon\sqrt{s})}$ for small  $r$ and $s$.
   The inverse Fourier transform gives
$P_{n,0}(s)=0$ for $n<0$. For   $n > 0$,  we close the contour along
the negative real axis.  After the substitution $s=-u^2$,  the $u$-integral can
be taken over the whole real axis, at the expense of a factor $\half$.
The integral is then Gaussian and leads to
  \BEQ
  \label{pn0t=}
P_{n,0}(t) \approx \frac{\epsilon n}{2\sqrt{\pi \va }(\va t-n)^{3/2}}\,
\exp\left[-\frac{\epsilon^2 n^2}{4\va (\va t-n)}\right] \quad {\rm for}
\quad 0\le n\le \va t \quad ,
\EEQ
which vanishes exponentially in $1/(\va t-n)$ as $n$ approaches $\va 
t$ from below
which shows that, for large $t$, all motors unbind  before they reach
the position $n = \va t$.

\section{Properties of the unbound motors}
Eventually all motors will become unbound and diffuse in the surrounding
solution. We now discuss how this behavior is affected by  the
filament. The FL  transformed distribution, $P_{ub}(q,r,s)$, of the 
unbound motors
  behaves as
  \BEQ
P_{ub}(q,r,s) \approx  \frac{4\epsilon\sqrt{s}}{(s-i\va 
r+\epsilon\sqrt{s})(q^2+4s)}
\label{FLTDistributionUnboundMotors}
\EEQ
for  small $q, r, s $ with $ q \sim \sqrt{s}$.
Using this asymptotic   expression , we can calculate the
  moments $N_a^\ub(t) \equiv \sum_{n}\sum_{m\neq 0}
n^a\,P_{n,m}$ for the  distribution of the unbound motors.
The fraction of unbound motors, $N_0^\ub$, satisfies
$N_0^\ub(t)=1-N_0(t)$.
The first moment $N_1^\ub$ determines the average displacement
and velocity,
$\bar n_{ub}$ and  $\bar v_{ub}$,  of the unbound motors
parallel to the filament via ${\overline n}_\ub \equiv N_1^\ub/N_0^\ub$
and ${\overline v}_\ub \equiv \d \bar n_\ub/\d t$.
For  small $t$,    ${\overline n}_\ub \approx
\frac{2}{3}\va t(1-\oq \epsilon\sqrt{\pi t})$ and ${\overline v}_\ub
\approx \frac{2}{3}\va (1-\frac{3}{8}\, \epsilon\sqrt{\pi t})$,
while for large $t$,
\BEQ
  \label{large_t_displ} {\overline
n}_\ub \approx \frac{2\va \,\sqrt{t}}{\epsilon\sqrt{\pi}}\,
[1-(\frac{\pi}{2}-1)\frac{1}{\epsilon\sqrt{\pi t}}]  \quad {\rm and} \quad
{\overline v}_\ub=\frac{\va }{\epsilon\,\sqrt{\pi t}}[1+{\cal
   O}(\frac{1}{t})].
\EEQ
Although each individual unbound motor has, on average, zero
  velocity, the  velocity
${\overline v}_\ub$ is non--zero,
since the cloud of unbound motors moves via repeated interactions with
the filament.
The average velocity of {\em both bound and unbound} motors is given by
$N_0 {\overline v}_b + N_0^\ub {\overline v}_\ub$. For large $t$, 
this latter quantity is
dominated by the unbound term    $N_0^\ub {\overline v}_\ub$ which becomes
asymptotically equal to
$  N_0 v_b$ in agreement with scaling arguments \cite{lipo173}.

For small and large $t$, the positional variance $\Delta n^2_\ub$ behaves as
\BEQ
\Delta n^2_\ub  \approx \va ^2t^2\frac{4}{45}\,(1-\frac{1}{8}\,
\frac{\epsilon\sqrt{t}}{\sqrt{\pi}})
\quad {\rm and} \quad
\Delta n_\ub^2 \approx {2(\pi-2)\va ^2t}/{\pi\eps^2} \quad ,
\EEQ
respectively.
The corresponding diffusion constant is given by
$D_\parallel (\infty) =(\pi-2)\va ^2/{\pi\eps^2}$.  The relation
$D_\parallel (\infty) \sim \va ^2/\eps^2$ implies that the parallel diffusion
in the bulk is strongly enhanced by the presence of the filament.

The   motor displacement  perpendicular to the filament
has the variance
\BEQ
\Delta m^2 =\frac{\epsilon}{2\pi} \int_0^\infty \d
s\,\frac{e^{-st}-1+st}{s^{3/2}(s+\epsilon^2)}
\EEQ
with the asymptotic behavior
\BEQ
\Delta m^2  \approx \frac{2\epsilon\,t^{3/2}}{3\sqrt\pi}  \quad
  {\rm and } \quad
\Delta m^2  \approx\half t\,(1-2\frac{1}{\epsilon\sqrt{\pi t}})
  \EEQ
for small and large $t$, respectively.  Thus,
the perpendicular diffusion is initially   suppressed
  but  the corresponding diffusion coefficient eventually
attains its unperturbed value $D_\perp (\infty) = \frac{1}{4}$.

In real space, the  probability distribution as given by
(\ref{FLTDistributionUnboundMotors}) becomes
\BEQ
\label{pnmt=} P_{n,m}(t)=\frac{\eps(\eps n+2|m|\va )} {2\sqrt{\pi \va }(\va
   t-n)^{3/2}}\, \exp\left[-\frac{(\eps n+2|m|\va )^2}{4\va (\va t-n)}
\right] \quad {\rm for} \quad
   0\le n\le \va t \quad (m\neq 0)  .
  \EEQ

\section{3--dimensional case ($d = 3$)}
Now consider the same kind of random walk but on a cubic lattice.
The jump probabilities per time step are
  $\os$ away from the filament,   $1-\gamma-\half
\delta-\frac{2}{3}\epsilon$  and $\half \delta$ in the forward and backward
direction, respectively, on the filament, and $\os\epsilon$ for each of the
four unbinding directions, where $\gamma$ is again the probability to
make no step at all.  The average   velocity within the bound state is $ \va
=1-\gamma-\delta-\frac{2}{3}\eps$, and the   probability  to stay in 
such a bound
state for
$t$ successive steps is
$\exp[-{2}\eps\,t/3]$.    We now denote the perpendicular coordinate by
$\mt=(m_1,m_2)$ and the perpendicular momentum by $\qb=(q_1,q_2)$.
After   integration over $q_1$ and $q_2$,  we obtain
\BEQ
  P_b(r,s)=\frac{3\,I(r,s)}{\epsilon+
   [3(1-\epsilon)s+ \half(\epsilon-3\delta)\,(e^{-ir}-1)
   -\half(6-6\gamma-3\delta-5\epsilon)(e^{ir}-1)] \,I(r,s)}
\label{Pbound3Dim}
\EEQ
with
  \BEQ
  I(r,s) \equiv \int_0^{2\pi}\frac{\d q_1}{2\pi}\,\frac{\d
   q_2}{2\pi}\, \frac{1}{3+3s-\cos r-\cos q_1-\cos q_2}
=\frac{\sqrt{m}}{\pi}\,K(m),\qquad
\EEQ
where $K(m) \equiv \int_0^{\pi/2}\d\phi/\sqrt{1-m\sin^2\phi}$ is a complete
elliptic integral at $m={4}/{(3+3s-\cos r)^2}$.

Using the explicit expressions for the FL transform of the probability
distributions,  one can now compute all quantities of interest for 
the 3--dimensional
case.  Since these  computations are somewhat tedious,
  we will only   present the results and describe  the details 
elsewhere~\cite{NKL}.

The statistical properties of the bound motors are   displayed in
Figs.~\ref{f.1} and \ref{f.2} where the
fraction $N_0$,  the average position  ${\overline n}_b$, and the  variance
$\Delta n^2_b$ are shown both for the  2-- and the 3--dimensional case.
For large $t$, the displacement of the unbound motors behaves as
$\overline{n}_\ub \approx \frac{3\va
   }{2\pi\eps}(\ln t+\gamma_E)$.  Again, this is equal to  the
asymptotic displacement averaged over both bound and unbound motors, and
the corresponding velocity   is given by $\va
N_0(t) \approx 3\va /({2\pi\epsilon t})$  as predicted by  scaling arguments
\cite{lipo173}.

The parallel variances of the bound and unbound motor positions
  exhibit    logarithmic corrections
   and  behave  as  $\Delta  n^2_b(t) \approx \frac{t}{3} +
\Psi(t)$ and $\Delta n^2_\ub (t) \approx \frac{t}{3} + \frac{1}{2} \Psi(t)$
for large $t$ with
$\Psi(t) \equiv \frac{9\va^2}{2\pi^2\eps^2}
\,[(\ln  t +\gamma_E)^2-\frac{\pi^2}{3}] $,
  $\tau_0 \equiv 3/16$,  and Euler's constant $\gamma_E \simeq 0.5778$.
The corresponding time--dependent diffusion coefficients
   behave  as
$  D_b (t) \approx  \frac{1}{6} + \frac{1}{2}\Psi^\prime(t)$
and  $  D_\parallel (t) \approx  \frac{1}{6} + \frac{1}{4}\Psi^\prime(t)$
with
$\Psi^\prime(t)  = \frac{9\va^2}{\pi^2\eps^2\,t}\, \,[\ln t
  +\gamma_E] $.
At the crossover time
$t\sim {1}/{\eps}$,  the logarithmic terms  are
$\sim {\ln(1/\eps)}/{\eps}$ and, thus,  are large compared to the value
$\frac{1}{6}$  which is  the diffusion coefficient of the motors in the absence
of the filament.
For   displacements in the directions perpendicular to the filament, one
finds the variances
$\Delta m^2_1 (t)= \Delta m^2_2 (t) \approx \frac{t}{3}-
\frac{1}{2\pi\eps}(\ln t+\gamma_E)$ for large $t$.

\section{Filament with several protofilaments}
Microtubules consist of 13 parallel protofilaments, on each of which
motors can run. To incorporate this in our model, we assume that the
pinning line at $m=0$ has $k$ internal states, and $k$ may be equal to
$13$. The average occupation of each of these states is denoted by
$p_{n,0}^{\,j}$, $j=1,..,k$.  There is a small probability,
$\half\zeta$, that a motor goes from protofilament $j$ to $j+1$ in one
time step, and similarly for going to protofilament $j-1$.  To take
into account the cylindrical structure of the microtubule, we identify
$j=k$ with $j=0$. We assume that, after unbinding, the motor may enter
the surrounding solution at a randomly chosen neighboring site of our lattice.

  In this model,  the
total fraction of motors at position $n$ along the tubule is
$P_{n,{\bf 0}}=\sum_{j=1}^k \, P_{n,{\bf 0}}^{\,j}$.  The motion on the
individual protofilaments can again be described by a master equation.
To solve it, we introduce the Fourier transform   $P_{n,{\bf
     0}}^\omega=\sum_{j=1}^k \, P_{n,{\bf 0}}^{\,j}\,e^{ij\omega}$  with
$\omega=2\pi \ell/k$ with $\ell=0,..,k-1$.  At the initial time $t=0$,
all motors are taken to be on the filament   position   $n=0$, ${\bf 
m}={\bf 0}$,
  and on the protofilament $j=0$.  We then find for the asymmetric modes with
$\omega\neq 0$
\BEQ
P_{b}^\omega(r,s)=\frac{1}{s+1-(1-\gamma-\half
   \delta- \frac{3}{2}\epsilon)e^{ir} -\gamma-\half\delta
   e^{-ir}+\zeta(1-\cos\omega)} \label{PboundOmega}
\EEQ
which, in this simple model,  has decoupled from $P_{b}^{\omega=0}(r,s) $.

  The bound fractions on the individual
protofilaments are obtained from (\ref{PboundOmega}) with  $r=0$
which leads to
\BEQ
  N_0^{\,j}(t) \equiv \sum_n p_{n,{\bf 0}}^{\,j}(t)
=\frac{1}{k}N_{ 0}(t) +\frac{1}{k}\sum_{\omega\neq 0}e^{-ij\omega}\,
e^{-\frac{3}{2}\epsilon t} e^{-\zeta(1-\cos\omega)t}
\EEQ
  The first term
describes the symmetric distribution of the motors over the $k$
protofilaments and decays algebraically, see Fig.~\ref{f.1}(a).    The second
term represents the asymmetry arising from the   initial
distribution.  There are two decay mechanisms of this
asymmetry.  The term $\exp(-\frac{3}{2}\epsilon t)$ expresses that
unbinding and subsequent rebinding to a randomly chosen
protofilaments restores the symmetry.   The factor
$\exp[-\zeta(1-\cos\omega)t]$ expresses that hopping to neighboring
protofilaments also restores the symmetry.  The asymmetry between the
average occupation of the various protofilaments thus disappears after
some transient time.

In summary, we have studied the walks of molecular
motors which arise from many diffusional encounters with a single filament.
In the absence of confining walls or boundaries, we have been able to
{\em analytically} calculate the  statistical properties of these walks
both for two and for  three dimensions.
The  analytical expressions were found to be in excellent agreement with
the results of MC simulations.
The diffusion of both bound and unbound motors was shown to be 
strongly enhanced
parallel  to the filament. This enhanced diffusion could be used, 
e.g., in order
to increase the rate of diffusion--controlled chemical reactions.

\end{document}